 \documentclass[final,5p,times,twocolumn]{elsarticle}

\usepackage{subfigure}
\usepackage{amssymb}
\usepackage{lipsum}
\usepackage{amsthm}
\usepackage{graphicx}
\usepackage{amsmath}
\usepackage[colorlinks=true, linkcolor=blue, citecolor=blue, urlcolor=blue]{hyperref}

\journal{Physics Letters B}

\begin{document}

\begin{frontmatter}



    \title{Kiselev black hole from semiclassical gravity with conformal anomaly }


\author[first,second]{Yu-Sen An\fnref{label1}}
\fntext[label1]{Corresponding author}
\ead{anyusen@nuaa.edu.cn}
\author[first]{Yu-Lin Chen}
\ead{15394885173@163.com}
\affiliation[first]{organization={Center for the Cross-disciplinary Research of Space Science and Quantum-technologies (CROSS-Q), College of Physics, Nanjing University of Aeronautics and Astronautics, Nanjing, 210016, China}}
\affiliation[second]{organization={MIIT Key Laboratory of Aerospace Information Materials and Physics,  Nanjing University of Aeronautics and Astronautics, Nanjing, 210016, China}}

\begin{abstract}
Investigating black hole physics in semi-classical gravitational theory is an interesting and important direction. When the backreaction of the conformal anomaly-induced effective stress-energy tensor is taken into account, the black hole spacetime can be largely altered which changes many properties of the black hole. In this work, rather than restricting our attention to the type-A anomaly, we incorporate the type-B anomaly and investigate how it modifies the spacetime structure obtained when only the type-A anomaly is considered. We find that for a specific relation between the central charges and the cosmological constant, the semiclassical Einstein equations can be solved analytically, yielding a metric that takes the exact form of the Kiselev black hole. After deriving the solution, we further discuss the properties of this new black hole solutions. Our work serves as an essential stepping stone for studying black hole geometries under the full conformal anomaly.  
\end{abstract}




\end{frontmatter}




\section{Introduction}
Black hole is a central prediction of general relativity which has been verified in various observations \cite{LIGOScientific:2016aoc,EventHorizonTelescope:2019dse,EventHorizonTelescope:2022wkp,EventHorizonTelescope:2022apq}. Besides a real astrophysical object, black hole is also an important platform to further investigate the fundamental question in theoretical physics. Since spacetime singularities are inevitable inside black holes\cite{Penrose:1964wq}, the general relativity- being a purely classical theory-cannot be the final theory describing the black hole physics which is only a low energy effective theory describing gravity. Moreover, general relativity and quantum physics are not consistent with each other which motivates physicists to develop a quantum theory of gravity. 

However, finding the quantum theory of gravity is a very challenging task since the gravity is not consistent with the framework of quantum field theory. The proper framework for formulating quantum gravity has been debated for half a century where many candidates can be chosen, such as string theory \cite{Schwarz:1982jn}, canonical quantum gravity \cite{DeWitt:1967yk,DeWitt:1967ub} and loop quantum gravity\cite{Ashtekar:2011ni}. Compared with these full theories of quantum gravity, semiclassical gravity is far more tractable and highly effective for capturing the quantum aspects of black holes \cite{DeWitt:1975ys,Hollands:2014eia}. Semi-classical gravity treats the spacetime as a classical aspect while quantize the matter field on the spacetime. This framework has led to major breakthroughs on various central questions regarding the black hole physics, such as the discovery of the Hawking radiation and Bekenstein entropy \cite{Hawking:1975vcx,Bekenstein:1973ur,Wald:1975kc}, establishment of the black hole thermodynamics \cite{Bardeen:1973gs} and recent progress toward resolving the black hole information paradox \cite{Almheiri:2019hni,Almheiri:2019psf,Penington:2019npb,Harlow:2014yka}. 

Among various quantum effects in curved spacetime, the conformal (or trace) anomaly plays a particularly crucial role \cite{Deser:1993yx,Duff:1993wm}. The conformal anomaly shows that a conformal field theory in flat spacetime is no longer conformal when placed on a curved background. The classically vanishing trace of stress-energy tensor $T^{\mu}_{\mu}=0$ due to conformal symmetry instead reads 
\begin{equation}\label{conformalanomaly}
    \langle T^{\mu}_{\mu}\rangle=c C^{\mu\nu\rho\sigma} C_{\mu\nu\rho\sigma}-aE_{(4)}
\end{equation}
where $C_{\mu\nu\rho\sigma}$ is the Weyl tensor and $E_{(4)}$ is the Gauss-Bonnet invariant $E_{(4)}=R^{\mu\nu\rho\sigma}R_{\mu\nu\rho\sigma}-4R_{\mu\nu}R^{\mu\nu}+R^{2}$. Therefore the conformal symmetry which is classically obeyed no longer exists in the quantum sense which is the essence of the anomaly. The first term in Eq.(\ref{conformalanomaly}) denotes the type-B anomaly and the last term denotes the type-A anomaly, the two constants $c$ and $a$ are two central charges which characterize the degrees of freedom of the underlying conformal field theory. 

By switching off the type-B anomaly and considering a typical equation of state $\rho=-p_{r}$, analytical solutions can be obtained which provide the simplest platform to investigate the effect of quantum anomaly on black hole physics \cite{Cai:2009ua,Cai:2014jea}. The analytical black hole solution is the same as the 4d Einstein-Gauss-Bonnet black hole up to a simple parameter shift \cite{Glavan:2019inb,Hu:2024ldp}. Various studies \cite{Hu:2024ldp,Chu:2025hes,RezaMehdizadeh:2025czq,Zi:2023qfk,An:2025xmb,Zhang:2026cqv,Yang:2025xck} investigate the thermodynamics and dynamics of this black hole solution and interesting phenomena occur. Afterwards, the generalization to the rotating case has also been found \cite{Fernandes:2023vux,Zhang:2023bzv}. Among the above developments, the most interesting property of this black hole solution is that by incorporating the type-A conformal anomaly, there exists isolated critical point \cite{Hu:2024ldp,Dolan:2014vba} where the second order phase transition will violate the scaling law which is universally obeyed in all the phase transition process \cite{Pelissetto:2000ek}. 

However, these previous findings rely on the simplifying assumption that the type-B conformal anomaly is neglected. To capture the backreaction of the full conformal anomaly, the type-B central charge cannot simply be set to zero. Therefore, in this work, we take both the type-A and type-B anomalies into account and search for exact analytical solutions to further clarify the relationship between semiclassical corrections and black hole geometry. The remainder of this paper is organized as follows. In Sec.\ref{bhsol}, we introduce the corrections of type-B anomaly on the modification of Einstein equation. In Sec.\ref{exact}, we show how to derive the new analytical black hole solution with both type-A and type-B conformal anomaly taken into account and demonstrate that it belongs to the Kiselev black hole family. In Sec.\ref{prop}, we analyze the thermodynamic and optical properties of this semiclassical black hole and compare them with those of the Schwarzschild–de Sitter geometry. Finally, we will conclude the paper and give some further outlooks in Sec.\ref{cond}. Throughout this work, the geometric unit $G=c=1$ is used.

\section{Backreaction effect of the type-B conformal anomaly: }\label{bhsol}
In this section, we first review the analytical black hole solution with only type A anomaly is taken into account.
By choosing a renormalization scheme in which the $\Box R$ contribution to the trace anomaly is absent, the resulting non-vanishing trace of stress-energy tensor reads
\begin{equation}
8\pi\langle T^\mu{}_{\mu}\rangle=BC^{\mu\nu\rho\sigma}C_{\mu\nu\rho\sigma}-AE_4 
\end{equation}
where we redefine the central charge to be
\begin{equation}
A=8\pi a ,\qquad B=8\pi c,\qquad \frac BA=\frac ca.
\end{equation}

In order to get the analytic black hole solution with spherical symmetry which reads
\begin{equation}
ds^2=-f(r) dt^2+\frac{dr^2}{f(r)}+r^2 d\Omega_2^2,
\end{equation}
the additional closure condition $\langle T^t{}_t\rangle=\langle T^r{}_r\rangle$ needs to be imposed which gives a physical constraint of energy density and radial pressure. 
Note that this closure condition is an additional model assumption in order to get the solution as the trace anomaly alone does not uniquely specify the full stress tensor of a general quantum state.  

By setting $u(r)=1-f(r)$, direct evaluation of the curvature invariants gives
\begin{align}\label{curvin}
R&=\frac{(r^2u)''}{r^2},\quad 
E_{(4)}=\frac{2(u^2)''}{r^2},\\&
C_{\mu\nu\rho\sigma}C^{\mu\nu\rho\sigma}=\frac{(r^2u''-2ru'+2u)^2}{3r^4},
\end{align}
By taking trace of the semi-classical Einstein equation
\begin{equation}
    R_{\mu\nu}-\frac{1}{2}g_{\mu\nu}R+\Lambda g_{\mu\nu}=8\pi G \langle T_{\mu\nu} \rangle
\end{equation}
we find 
\begin{equation}\label{traceeq}
    R-4\Lambda=AE_{(4)}-B C^{\mu\nu\rho\sigma}C_{\mu\nu\rho\sigma}
\end{equation}
Substituting the spherical symmetrical black hole ansatz and the curvature invariant expressions Eq.(\ref{curvin}) into Eq.(\ref{traceeq}), the above Einstein equation reduces to
\begin{equation}\label{eq:master}
(r^2u-2Au^2)''=4\Lambda r^{2}-\frac{B}{3r^2}(r^2u''-2ru'+2u)^2.
\end{equation}

For $B=0$, integrating Eq.~\eqref{eq:master} twice yields
\begin{equation}
r^2u_A-2Au_A^2=-\frac{r^4}{\ell^2}+2Mr-Q^2.
\end{equation}
where $M$ and $Q$ are two integrable constants and $l=\sqrt{-\frac{3}{\Lambda}}$. 
Choosing the branch continuously connected to Einstein gravity gives the analytical solution as obtained in \cite{Cai:2009ua,Cai:2014jea}
\begin{equation}\label{eq:Cai}
f_A(r)=1-\frac{r^2}{4A}
\left[1-\sqrt{1+\frac{8A}{\ell^2}-\frac{16AM}{r^3}+\frac{8AQ^2}{r^4}}\right].
\end{equation}
which is an asymptotically AdS black hole with only type-A conformal anomaly correction. 

However with the type-B conformal anomaly taken into account, the Einstein equation can not be easily integrated which makes it considerably more difficult to solve.  For the most general case without any relation among the parameters $A$, $B$ and $\Lambda$, the Einstein equation can only be solved in terms of series expansion \cite{Fernandes:2023vux,Borah:2025crf,Tsujikawa:2023egy}. While series expansions provide valuable insights, the lack of an exact closed-form expression obscures many global properties of the black hole. Thus, it remains an interesting open question whether exact analytical solutions exist when both the type-A and type-B anomalies are non-vanishing. 
\section{Exact black hole solution with full conformal anomaly:}\label{exact}
In this section, we turn on the type-B anomaly correction and try to find an analytic solution when both anomaly terms coexist. 

Firstly, considering the trial ansatz of $u(r)$ to be 
\begin{equation}\label{testfunc}
    u(r)=kr^{2}+C r^{p}
\end{equation}
where $k$ encodes the de-Sitter radius $k=1/l_{ds}^{2}$ and plugging this into the Eq.(\ref{eq:master}), we get 
\begin{equation}
\begin{aligned}
    &4(3k-6Ak^{2}-\Lambda) r^{2}+C(1-4Ak)(p+1)(p+2)r^{p}+\\&C^{2}(-4Ap(2p-1)+\frac{B}{3}(p-1)^{2}(p-2)^{2})r^{2p-2}=0
    \end{aligned}
\end{equation}
For this equation to hold, the coefficients of each power of $r$ must vanish, which yields the conditions
\begin{equation}
    3k-6A k^{2}-\Lambda=0
\end{equation}
\begin{equation}
    1-4Ak=0
\end{equation}
This gives a simple relation between the central charge and the cosmological constant. 
\begin{equation}\label{asumpp}
    k=\frac{1}{4A}, \quad  \Lambda=\frac{3}{8A}
\end{equation}
Moreover, the vanishing coefficient of the $r^{2p-2}$ term gives a relation between two central charges 
\begin{equation}
    \frac{B}{A}=\frac{12p(2p-1)}{(p-1)^{2}(p-2)^{2}}
\end{equation}
For p taking different values, we get different exact black hole solution with typical relation among central charges. 

For example, if $p=-1$, we find $A=B$ and the solution is the same as Schwarzschild de-Sitter spacetime. For this case, the metric function takes the form 
\begin{equation}
    f(r)=1-\frac{r^{2}}{4A}-\frac{C}{r}
\end{equation}

While for $p=-2$, on the other hand, the metric function reduces to 
\begin{equation}
    f(r)=1-\frac{r^{2}}{4A}-\frac{C}{r^{2}}
\end{equation}
For this case, the solution is the Reissner-Nordstrom-de Sitter black hole with zero ADM mass. For $C<0$, the solution is a naked singularity while for $C>0$, there is a horizon around the singularity at the price of violating standard energy conditions. 

The main focus of this work is the case of equal central charges $A=B$ \footnote{$p=-1$ condition must lead to $A=B$ but not vice versa} where particularly rich behavior emerges. In this case,  the equation 
\begin{equation}
    12p(2p-1)=(p-1)^{2}(p-2)^{2}
\end{equation}
yields the roots $p_{0}=-1$ and $p_{\pm}=4\pm 2\sqrt{3}$, which actually leads to more structures of the black hole solution besides the Schwarzschild de-Sitter solution. However, due to the constraint of the trial ansatz Eq.(\ref{testfunc}), it is obscure whether the metric can take the form with both polynomials $u(r)=kr^{2}+\frac{\mu}{r}+C r^{p_{\pm}}$. However, for the special case where Eq.(\ref{asumpp}) and $A=B$ are both satisfied, we can relax the ansatz in Eq.(\ref{testfunc}) to obtain the more general analytical solution.\footnote{Note that as Eq.(\ref{asumpp}) must hold for our new analytical solution, the cosmological constant $\Lambda$ must be positive since the central charge $A$ is positive. This scenario is very different from the case in Ref.\cite{Cai:2009ua,Cai:2014jea}. }

For this special case, it can be seen that an algebraic simplification of the Einstein equation appears. If we set $u(r)=\frac{r^{2}}{4A}+v(r)$, the Einstein equation Eq.(\ref{eq:master}) simplifies to be 
\begin{equation}
    (r^{2}v''-2rv'+2v)^{2}-6r^{2}(v^{2})''=0
\end{equation}
This equation can be factorized to be 
\begin{equation}
\begin{aligned}
    &[r^{2}v''-(2+2\sqrt{3})rv'-(4+2\sqrt{3})v] \times \\& [r^{2}v''-(2-2\sqrt{3})rv'-(4-2\sqrt{3})v]=0
    \end{aligned}
\end{equation}
which means that the condition of satisfying Einstein equation is broken down to either of the following two equations hold 
\begin{equation}
    r^{2}v''-(2+2\sqrt{3})rv'-(4+2\sqrt{3})v=0
\end{equation}
\begin{equation}
    r^{2}v''-(2-2\sqrt{3})rv'-(4-2\sqrt{3})v=0
\end{equation}
The above ordinary differential equations are easily solved which leads to the following two kinds of exact solution 
\begin{equation}\label{rpsol}
    f(r)=1-\frac{r^{2}}{4A}-\frac{\mu}{r}-Cr^{4+2\sqrt{3}}
\end{equation}
\begin{equation}
    f(r)=1-\frac{r^{2}}{4A}-\frac{\mu}{r}-Cr^{4-2\sqrt{3}}
\end{equation}
This is the analytical black hole solution with both type-B and type-A conformal anomaly effect are taken into account. 

For the case of Eq.(\ref{rpsol}), the asymptotic behavior of the geometry is changed since $p_{+}=4+2\sqrt{3}>2$. This will leads to the big-rip behavior whose effective stress energy tensor is of phantom type \cite{Caldwell:2003vq}. We do not consider this situation in this work. 

Thus we will mainly focus on the case $p_{-}=4-2\sqrt{3}$ as the asymptotic geometry is still of de-Sitter type. In Ref.\cite{Kiselev:2002dx},  Kiselev investigated the effect of quintessence dark energy on the black hole geometry. Despite some critiques on the terminology \cite{Visser:2019brz}, the Kiselev solution remains an interesting platform to investigate the relation between dark energy and black hole physics which receives many attentions \cite{Ghosh:2015ovj,Zeng:2020vsj,Chen:2005qh,Li:2014ixn,Xu:2016jod}. In the presence of both quintessence and positive cosmological constant, the black hole solution takes the following form 
\begin{equation}
    f(r)=1-\frac{\mu}{r}-\frac{\Lambda r^{2}}{3}-\frac{K}{r^{3w+1}}
\end{equation}
where $w$ is the state parameter which lies in the region $-1<w<-1/3$. For the solution we obtained, we can find that $p_{-}=4-2\sqrt{3}$ corresponds to $w=-\frac{5-2\sqrt{3}}{3}\approx -0.512$ 
which lies in the region given by Kiselev \cite{Kiselev:2002dx}. Thus the new semiclassical black hole solution with both type-B and type-A conformal anomaly corrections is of the Kiselev type. For the Kiselev black hole solution, effective energy density and pressure are directly imposed without further microscopic origin, however in our case, we give a concrete semiclassical mechanism for generating this kind of black hole solution. 

\section{Properties of the new conformal anomaly corrected black hole solution} \label{prop}
In this section, we turn to investigate the properties of the black hole solution where the blackening factor is $f(r)=1-\frac{r^{2}}{4A}-\frac{\mu}{r}-Cr^{4-2\sqrt{3}}$. From the metric function, the effective cosmological constant can be read off as 
\begin{equation}
    \frac{\Lambda_{eff}}{3}=k=\frac{1}{4A}, \quad \Lambda_{eff}=\frac{3}{4A}=2\Lambda
\end{equation}
This is different from the classical case since the de-Sitter radius is related to the effective cosmological constant $l_{dS}^{2}=\frac{3}{\Lambda_{eff}}$ instead of the bare cosmological constant $\Lambda$ in the equations of motion. 

For the physical meaning of $C$, this is related to the strength of the conformal anomaly. For this background, the conformal anomaly term reads
\begin{equation}
    \langle T^{\mu}_{\mu}\rangle_{C}=-\frac{3}{16\pi G A}-\frac{(5-2\sqrt{3})(6-2\sqrt{3})}{8\pi G} C r^{2-2\sqrt{3}}.
\end{equation}
The horizon position for this black hole background can be calculated as 
\begin{equation}\label{Crit}
    f(r_{h})=0 \to C=\frac{1-\frac{r_{h}^{2}}{4A}-\frac{\mu}{r_{h}}}{r_{h}^{4-2\sqrt{3}}}
\end{equation}
which gives a relation between physical parameter $C$ and horizon radius $r_{h}$. As the derivative of $C(r)$ reads
\begin{equation}
    C'(r)=\frac{(5-2\sqrt{3})\mu-(4-2\sqrt{3})r-\frac{2\sqrt{3}-2}{4A}r^{3}}{r^{6-2\sqrt{3}}},
\end{equation}
it starts from positive value and then decrease to be negative. This means that the function $C(r)$ can only have one maximum. 

The maximum corresponds to $C'(r)=0$ which is 
\begin{equation}
    (2\sqrt{3}-2) r_{N}^{3}+4(4-2\sqrt{3})A r_{N}-4A(5-2\sqrt{3})\mu=0, 
\end{equation}
thus the corresponding parameter $C$ is 
\begin{equation}
    C_{crit}=\frac{1-\frac{3r_{N}^{2}}{4A}}{(5-2\sqrt{3})r_{N}^{4-2\sqrt{3}}}
\end{equation}

Consequently, for $C < C_{\text{crit}}$, Eq.(\ref{Crit}) has two real roots which means the spacetime possesses two horizons where the larger root corresponds to the cosmological horizon and the smaller root corresponds to the black hole horizon. As $C$ increases, which corresponds to an enhanced trace of the stress-energy tensor, the black hole horizon expands outward while the cosmological horizon shrinks inward. 

At $C = C_{\text{crit}}$, Eq.(\ref{Crit}) has only one root, thus the original two horizons coincide with each other, reaching the Nariai limit. For $C > C_{\text{crit}}$, no horizons exist. Therefore, for the spacetime to admit two horizons, the parameter $C$ cannot take arbitrary values. 

\begin{figure}[h!]
    \centering
    \includegraphics[width=0.5\textwidth]{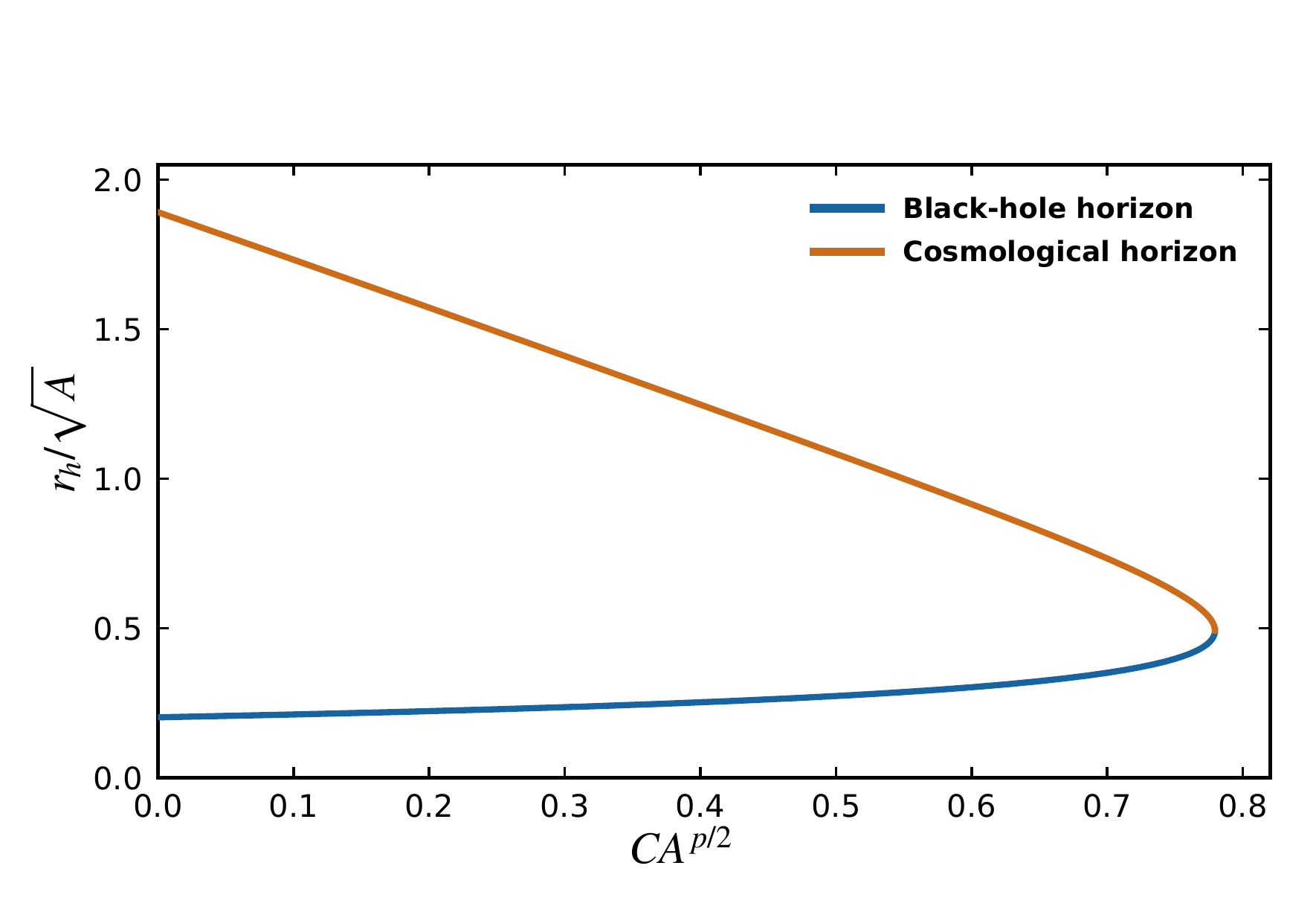}
    \caption{The relation between black hole horizon and cosmological horizon with respect to parameter $C$ for the Kiselev black hole with $p=4-2\sqrt{3}$. We rescale the parameter to be $r_{h} \to r_{h}/\sqrt{A}$ and $C \to CA^{p/2}$ and fix the parameter to be $\mu/\sqrt{A}=0.2$. }
    \label{horizon}
\end{figure}

For the small $C$ case, the correction of horizon radius can be 
\begin{equation}
    r_{b,c}=r_{b0,c0}+\frac{Cr_{b0,c0}^{5-2\sqrt{3}}}{1-\frac{3r_{b0,c0}^{2}}{4A}}+O(C^{2})
\end{equation}
where $r_{b0}$ and $r_{c0}$ are the horizon positions of Schwarzschild de-Sitter black hole. We plot the correction of the two horizon radius in the Fig.\ref{horizon}. The presence of anomaly will increase the black hole horizon radius while decrease the cosmological horizon radius until these two horizons coincide with each other which corresponds to the Nariai limit. 

As there are two horizons with different temperature, the whole system is not in thermal equilibrium. Traditional Euclidean approach \cite{Gibbons:1976ue,Draper:2022xzl} to the black hole thermodynamics fails due to the existence of conical singularity. So in this work, instead of establishing the thermodynamics of this black hole, we only demonstrate the temperature correction of this background due to conformal anomaly. 

\begin{figure}[h!]
    \centering
    \includegraphics[width=0.5\textwidth]{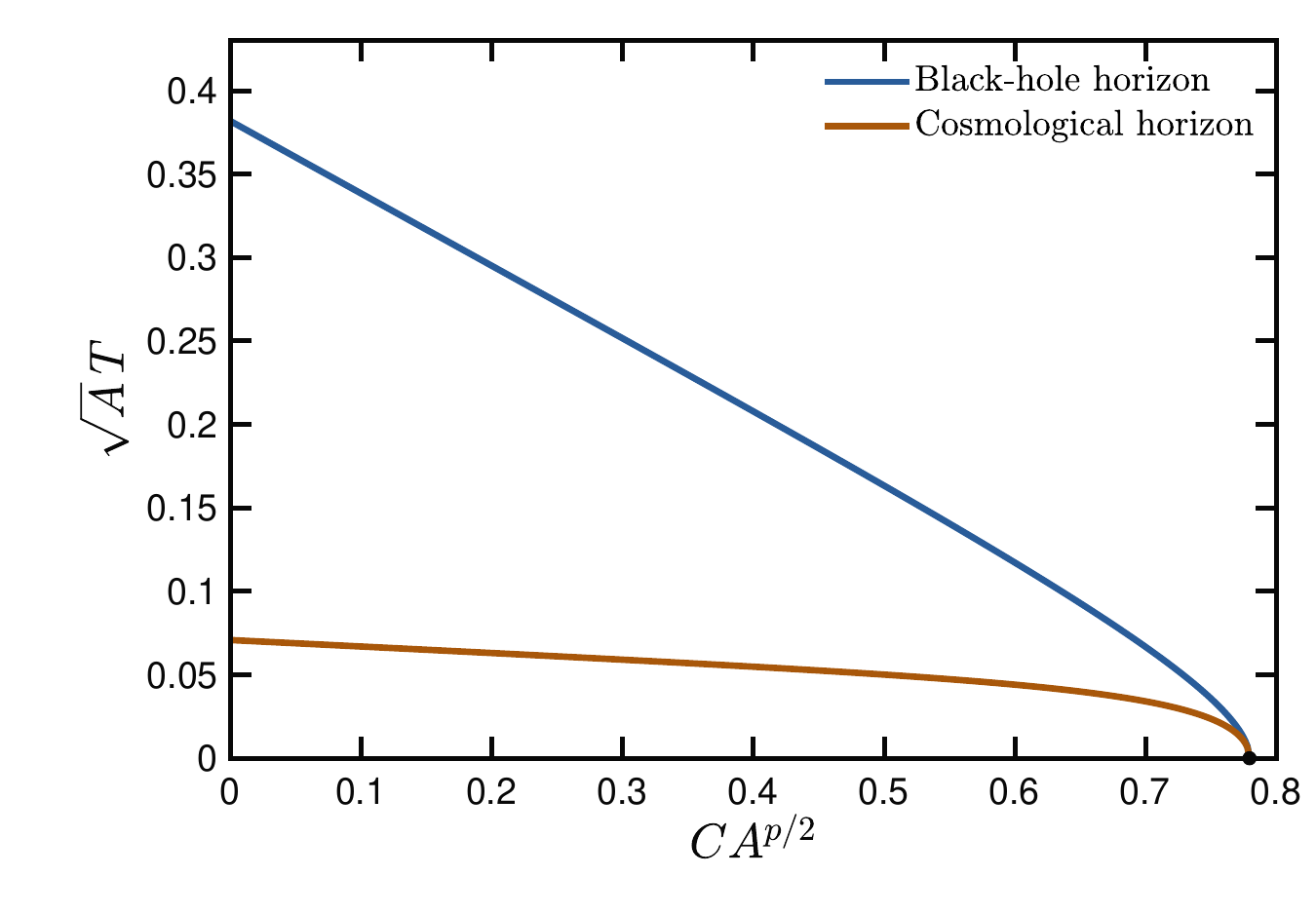}
    \caption{Temperatures of the two horizons where we choose $p=4-2\sqrt{3}$ and $\mu/\sqrt{A}=0.2$. The parameters have been rescaled to be dimensionless. It can be seen that the  surface gravity (temperature) of both horizons reduce to zero in the Nariai limit.  }
    \label{horizon_temperature}
\end{figure}

The temperature associated to the two respective horizons read
\begin{equation}
    T_{b}=\frac{\kappa_{b}}{2\pi}=\frac{f'(r_{b})}{4\pi}, \quad T_{c}=\frac{\kappa_{c}}{2\pi}=-\frac{f'(r_{c})}{4\pi}.
\end{equation}
By directly plugging in the expression of the blackening factor $f(r)$, the temperatures of the two horizons are 
\begin{equation}
    T_{b}=\frac{1}{4\pi}[\frac{1}{r_{b}}-\frac{3r_{b}}{4A}-(5-2\sqrt{3})C r_{b}^{3-2\sqrt{3}}],
\end{equation}
\begin{equation}
    T_{c}=\frac{1}{4\pi}[-\frac{1}{r_{c}}+\frac{3r_{c}}{4A}+(5-2\sqrt{3})C r_{c}^{3-2\sqrt{3}}].
\end{equation}
The effect of anomaly parameter on the corresponding horizon temperatures has been shown in Fig.\ref{horizon_temperature}. The temperature monotonically decrease and finally reduce to zero in the Nariai limit. 
It should be noted that Fig.\ref{horizon_temperature} does not mean that the Nariai black hole's temperature is zero since the temperature also depends on the actual observers we choose. If we first calculate the temperature for the static observer (with zero acceleration) and then take the Nariai limit, the resulting temperature will be nonzero. According to the Bousso-Hawking proposal in Ref.\cite{Bousso:1996au}, the temperature of the local static observer is $T=\frac{f'(r_{h})}{4\pi\sqrt{f(r_{0})}}$ where $r_{0}$ satisfies $f'(r_{0})=0$. When the Nariai limit is approached, Bousso-Hawking temperature will take the form 
\begin{equation}
    T_{N}=\frac{1}{2\pi}\sqrt{-\frac{f''(r_{N})}{2}}
\end{equation}
In the Nariai limit of our background, the critical mass and anomaly parameters read 
\begin{equation}
    \mu_{crit}=\frac{r_{N}}{5-2\sqrt{3}} (4-2\sqrt{3}+(2\sqrt{3}-2)\frac{r_{N}^{2}}{4A}), \quad C_{crit}=\frac{1-\frac{3r_{N}^{2}}{4A}}{(5-2\sqrt{3})r_{N}^{4-2\sqrt{3}}}
\end{equation}
by plugging these expressions into the second order derivatives of the $f(r_{N})$, the limiting temperature reads
\begin{equation}\label{Ntemp}
    T_{N}=\frac{1}{2\pi}\sqrt{\frac{3(\sqrt{3}-1)}{4A}+\frac{2-\sqrt{3}}{r_{N}^{2}}}
\end{equation}
which is non-zero and depends on the central charge $A$. In the Schwarzschild de-Sitter limit where $C\to 0$, the Nariai radius is $r^{2}_{N}=\frac{4A}{3}$. The corresponding Nariai temperature is 
\begin{equation}
    T_{N}=\frac{\sqrt{3}}{4\pi \sqrt{A}}=\frac{\sqrt{\Lambda_{eff}}}{2\pi}
\end{equation}
which means that the result Eq.(\ref{Ntemp}) is consistent with the Nariai temperature of Schwarzschild de-Sitter black hole \cite{Bousso:1996au,Fernando:2013mex}. 

Furthermore, the dynamical properties of the black hole should also be discussed. The most prominent behavior is to investigate the behavior of the photon sphere and black hole shadow. 

For massless particles moving in the new background where $ds^{2}=g_{\mu\nu}\dot{x}^{\mu}\dot{x}^{\nu}=0$, the Lagrangian is given by 
\begin{equation}
    L=\frac{1}{2}(-f(r)\dot{t}^{2}+\frac{\dot{r}^{2}}{f(r)}+r^{2}\dot{\theta}^{2}+r^{2}\sin^{2}\theta \dot{\phi}^{2})
\end{equation}
where the dot is with respect to the affine parameter. 

Due to static property and spherical symmetry, we can choose the particles to lie on the equatorial plane and two conserved quantity can be given 
\begin{equation}
    E=-f(r)\dot{t}, \quad L=r^2\dot{\phi}
\end{equation}
So the radial motion is 
\begin{equation}\label{partradial}
\dot{r}^{\,2}+V_{eff}(r)=E^2
\end{equation}
where the effective potential is defined as $V_{\mathrm{eff}}(r)=L^2\frac{f(r)}{r^2}=\frac{L^{2}}{r^{2}}(1-\frac{r^{2}}{4A}-\frac{\mu}{r}-Cr^{4-2\sqrt{3}})$. 

The circular orbit must satisfies $\dot{r}=\ddot{r}=0$, thus by taking derivatives of Eq. (\ref{partradial}), the circular orbit condition is 
\begin{equation}
    V_{eff}'(r_{ph})=0 \to rf'(r)-2f(r)=0
\end{equation}
Thus our photon sphere radius satisfies following equation 
\begin{equation}
   2r_{\mathrm{ph}}=3\mu+\left(2\sqrt{3}-2\right)C r_{\mathrm{ph}}^{\,5-2\sqrt{3}} 
\end{equation}
The stability of the photon sphere can be deduced by the sign of $V''(r_{ph})$, where 
\begin{equation}
    V_{eff}''(r_{ph})=L^{2} \frac{rf''(r)-f'(r)}{r^{3}} |_{r=r_{ph}}
\end{equation}
It can be easily found that $V_{eff}''(r_{ph})<0$ if $r_{b}<r_{ph}<r_{c}$ which means the photon sphere is unstable. 

\begin{figure}[h!]
    \centering
    \includegraphics[width=0.5\textwidth]{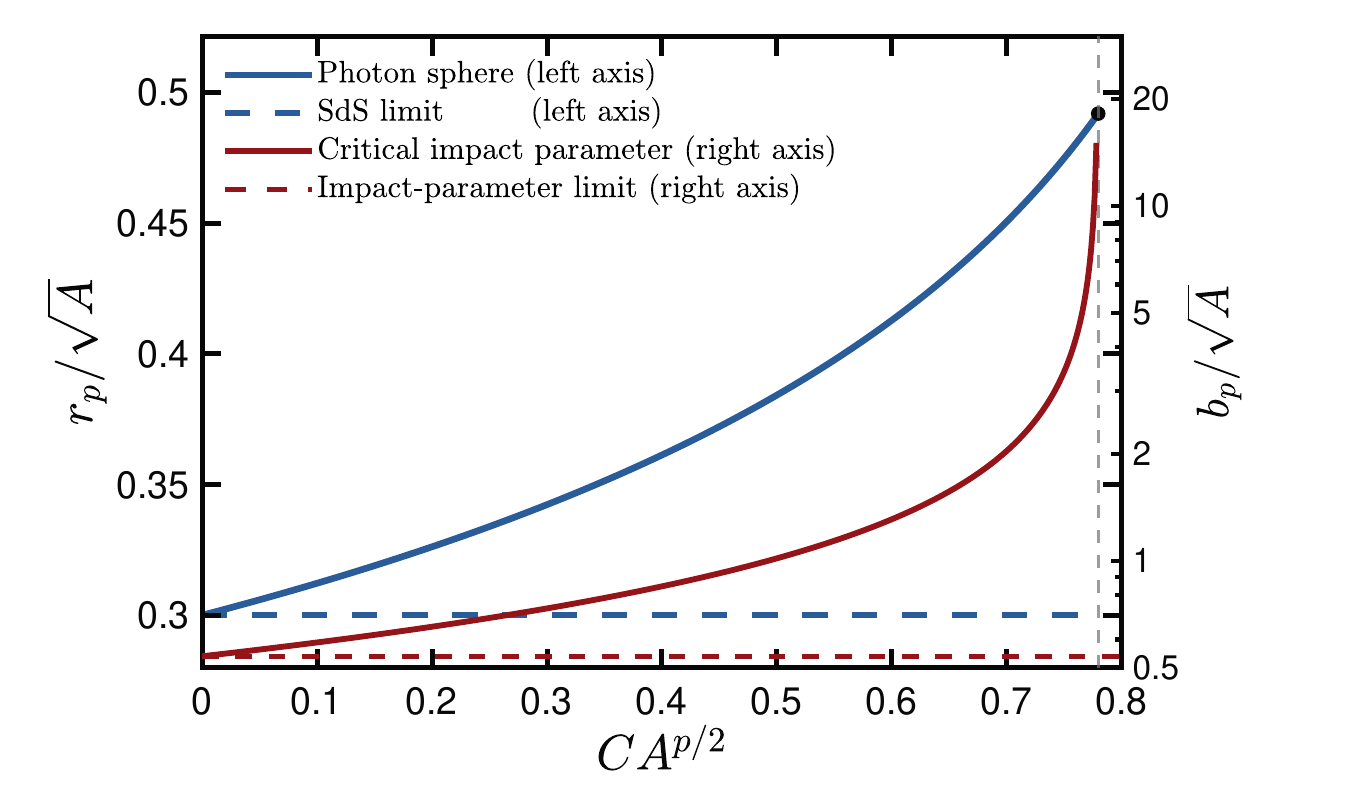}
    \caption{The relation between photon sphere radius and critical impact parameter with the anomaly parameter $C$. The photon sphere radius and critical impact parameter monotonically increase as we increase anomaly parameter while approaching the Schwarzschild de-Sitter limit in the limit $C\to 0$. We set $p=4-2\sqrt{3}$ and $\mu/\sqrt{A}=0.2$ through the numerical computation and the parameters have been rescaled to maintain dimensionless. }
    \label{ps}
\end{figure}

By numerically solving the above equation, we plot the relation between the photon sphere radius and anomaly parameter $C$ in Fig.\ref{ps}.
Starting from the photon sphere of Schwarzschild de-Sitter black hole where $r_{p}=\frac{3}{2}\mu=3M$, the radius monotonically increase as we increase the anomaly parameter $C$. Moreover, the shadow radius is determined by the critical impact parameter $b_{p}$ where 
\begin{equation}
    b^{2}_{p}=\frac{r^{2}_{p}}{f(r_{p})}=\frac{3r_{p}^{2}}{1-\frac{3r_{p}^{2}}{4A}-(5-2\sqrt{3})Cr_{p}^{4-2\sqrt{3}}}
\end{equation}
we also plot the relation between $b_{p}$ and $C$ in Fig.\ref{ps}. We see that for fixed $\mu$ and $A$, the shadow radius will also monotonically increase as we increase the anomaly parameter $C$ thus apparently, the semiclassical quantum corrections will enlarge the black hole shadows. 

The apparent divergence of $b_{p}$ in Fig.\ref{ps} in the Nariai limit is due to the fact that the energy is not formally normalized. As critical impact parameter is related to the energy by $b_{p}=L/E$, it relies on the choice of observer. If we  use the coordinate time $t$ in the computation, there will be divergence in Nariai limit which also appears in Schwarzschild-de Sitter case. However, according to Ref.\cite{Bousso:1996au}, if we choose the static observer at $r_{0}$ whose energy is $E_{loc}=E/\sqrt{f(r_0)}$ to compute the critical impact parameter, there will be no divergence. 

\section{Conclusion and Discussion} \label{cond}
In this work, we have obtained a Kiselev-like black hole solution in semiclassical gravity with conformal anomaly corrections. Rather than switching off the type-B conformal anomaly, we retain both the type-A and type-B anomalies and derive an exact analytical solution for the specific case $A=B$ and $\Lambda=\frac{3}{8A}$. This solution is novel since this incorporates the effect of both types of conformal anomaly correction which serves as an interesting new theoretical laboratory for investigating black hole physics with semiclassical corrections. 

Many further directions can be pursued in the future. Firstly we only consider the spherically symmetric black hole solution in this work. We can seek for the existence of analytical rotating black hole solutions in the future research. Moreover, we can calculate more properties of the black hole solution we found such as its stability condition under certain perturbations and particle motion behaviors around it. These studies will provide observational methods for detecting this kind of anomaly correction in actual observations. 
\section*{Acknowledgements}
This work is supported by the National Natural Science Foundation of China (NSFC) under Grants No.12405066. YSA is also supported by the Natural Science Foundation of Jiangsu Province under Grant No. BK20241376 and NUAA funding (No. 1018-ILF26028). 


\bibliographystyle{elsarticle-num} 
\biboptions{sort&compress}
\bibliography{msc}

\end{document}